# Restoring Pristine Bi$_2$Se$_3$ Surface with an Effective Se Decapping Process

Jixia Dai, Wenbo Wang, Matthew Brahlek, Nikesh Koirala, Maryam Salehi, Seongshik Oh, and Weida Wu[†]

Department of Physics and Astronomy, Rutgers University, Piscataway NJ 08854, USA

**Abstract**

High quality thin films of topological insulators (TI) such as Bi$_2$Se$_3$ have been successfully synthesized by molecular beam epitaxy (MBE). Although the surface of MBE films can be protected by capping with inert materials such as amorphous Se, restoring an atomically clean pristine surface after decapping has never been demonstrated, which prevents in-depth investigations of the intrinsic properties of TI thin films with ex-situ tools. Using high resolution scanning tunneling microscopy/spectroscopy (STM/STS), we demonstrate a simple and highly reproducible Se decapping method that allows recovery of the pristine surface of extremely high quality Bi$_2$Se$_3$ thin films grown and capped with Se in a separate MBE system then exposed to atmosphere during transfer into the STM system. The crucial step of our decapping process is the removal of the surface contaminants on top of amorphous Se before thermal desorption of Se at a mild temperature (~210 °C). This effective Se decapping process opens up the possibility of *ex-situ* characterizations of pristine surfaces of interesting selenide materials and beyond using cutting-edge techniques.

**Introduction --** Topological insulators (TIs) are celebrated for their insulating bulk and metallic surfaces, with the surface states protected by the topology of their electronic structure. [1-3] Although these surface states are immune to backscattering, [4] disordered or dirty surfaces prohibit systematic studies of their properties. [5,6] For Bi$_2$Se$_3$, a well-established large gap TI, [7,8] contaminated surfaces have been shown to affect their transport performance, [5,9] and it is especially detrimental for surface sensitive probes such as scanning tunneling microscopy (STM) and angle-resolved photoelectron spectroscopy (ARPES), for which atomically clean surfaces are necessary for reliable measurements. Therefore, in order to study the intrinsic properties of Bi$_2$Se$_3$ thin films and other TI samples grown by the state-of-the-art molecular beam epitaxy (MBE) with STM, a natural way is to integrate these two sophisticated instruments into one system which allows for in-situ characterization. [10,11] However, as one tries to incorporate more and more modern experimental techniques, the complexity and cost of such systems goes up, and becomes impractical. Moreover, the stability and resolution of individual instruments (such as ARPES or STM) will be ultimately limited by other parts of the combined system. Alternatively, it is possible, yet inconvenient, to transfer samples without breaking vacuum using a dedicate vacuum suitcase system. [12] A much simpler method is to protect the sample surface with a capping layer, which can be removed and thereby restoring the pristine surface when needed.

† Address correspondence to wdwu@physics.rutgers.edu

Due to the relatively low sublimation temperature of selenium in high vacuum, a technique involving Se caping and thermal decapping has been used to protect MBE films of selenides such as ZnSe and $CuInSe_2$. [13-15] Recently it has also been applied to $Bi_2Se_3$ and FeSe thin films for ex-situ studies, e.g. ARPES. [16-19] Thus far, there has not been any STM study on surfaces of $Bi_2Se_3$ thin films prepared by this method. Since STM/STS has been one of the major surface sensitive techniques to study the physics of the topological surface states in TI's, [4,11,20-23] it is therefore imperative to perform high resolution STM/STS studies of the $Bi_2Se_3$ surfaces after Se decapping.

**Results and Discussions**

Using a low temperature STM (Omicron), we performed a systematic study of various decapping processes of $Bi_2Se_3$ thin films synthesized in a separate ex-situ MBE chamber. Although it is generally believed that the surfaces of chalcogenides are relatively inert [24,25], there is always some surface adsorption in ambient condition. This is demonstrated by STM images of $Bi_2Se_3$ films without a Se capping layer that were directly transferred into the load-lock of our STM system after a brief exposure to air (~10 minutes). After loading the sample into the ultra-high vacuum (UHV) chambers, it was directly inserted into the STM head and cooled to 48 K. As shown in the STM image in Figure 1, the sample surface shows large terraces of a few hundred nanometers and straight step edges, indicating the high sample quality. The step height is about 9.5 Å as shown in the line profile in Fig. 1b, corresponding to the thickness of one $Bi_2Se_3$ quintuple layer (QL). However, a closer examination of the terraces reveals two types of surfaces, one being flat and the other one being rough at the atomic scale, as illustrated in Figs. 1b and 1d. The rough regions cover ~70% of the surface with an approximate thickness ~2-3 Å. This is likely due to adsorption of atmospheric contaminants during sample transfer. Interestingly, the rough region is not completely disordered, showing a "partial" order similar to the sematic liquid crystal phase [26], as shown in Fig. 1d. In contrast, high resolution STM image (Fig. 1e) of the atomically flat region reveals a well ordered triangular lattice with lattice constant of ~4 Å, indicating that the flat region is pristine $Bi_2Se_3$ and the atoms in Fig. 1d are the top layer Se. In short, a brief exposure to air (~10 minutes) would result in 70% of the $Bi_2Se_3$ surface covered with a contamination layer. Therefore, it is necessary to cover the surface of $Bi_2Se_3$ thin films if exposure to air is unavoidable for ex-situ characterization or device fabrication. Naturally, amorphous Se is chosen to cap the surface of $Bi_2Se_3$.

After the growth of the $Bi_2Se_3$ films, a thick layer (100 nm) of amorphous Se film was deposited on top of the Bi$_2$Se$_3$ thin film. Visually the Se capping layer changes the film color from silvery to shiny brown or blue. Figure 2a shows an atomic force microscopy (AFM) image of the typical morphology of the Se capping layer, which is quite rough (roughness ~2 nm as shown in Fig. 2b and Fig. S1 in the Electronic Supplementary Material (ESM)) compared to the uncapped films. [17] Also, the step edges of the Bi$_2$Se$_3$ film are invisible after Se capping. These Se capped samples were introduced into the STM chamber after exposed to air for a few hours or longer. To remove the Se capping layer for STM studies, we first follow the typical decapping method employed by many groups, i.e. direct thermal desorption of amorphous Se. [16,17,27] We have systematically studied the films obtained with a wide range of desorption temperature $T_d$ (see ESM Fig. S2 for details).

Figure 2c shows an STM image of Bi$_2$Se$_3$ film after direct thermal desorption at $T_d$ ~ 210 °C. (See ESM Fig. S2 for more data). The film morphology is similar to that of the uncapped film (Fig. 1a), indicating that the Se capping layer was removed with a relatively low $T_d$. However, a closer examination of a smaller area (Fig. 2c inset) reveals that areas of clean Bi$_2$Se$_3$ only consist of a small fraction the total area, while a large portion of the surface is still covered by a nanometer scale contamination layer. Differential conductance (dI/dV) measurement on clean surface that is away from the contaminants (Fig. 2d) shows typical Bi$_2$Se$_3$ spectra with the Dirac point at -210 meV, in good agreement with previous in-situ STS measurements on MBE films. [22] No significant reduction of surface contaminants is observed with repeated thermal desorption up to $T_d$ ~ 350 °C (see ESM Fig. S2), indicating they have much higher binding energy with Bi$_2$Se$_3$ surface than amorphous Se. Therefore, our results suggest that the optimal $T_d$ for decapping amorphous Se layer on Bi$_2$Se$_3$ is ~ 200-300 °C. On the other hand, excessively high $T_d$ leads to a qualitative change of film morphology as shown in Fig. 2e (more results in ESM Fig. S3), indicating a decomposition of surface Bi$_2$Se$_3$ QLs. Interestingly, this film morphology suggests that likely the grain boundaries and dislocations (Fig. 2e and ESM Fig. S3) are nucleation regions for decomposition. Accelerated decomposition around the dislocations is strikingly similar to the growth process of Bi$_2$Se$_3$ thin films. [28] Note that the optimal growth temperature ($T_g$) of Bi$_2$Se$_3$ films is ~ 300 °C, which is consistent with our observation of film decomposition at $T_d$ >350 °C. The degradation of the film at high temperature is further corroborated by STS measurements on the clean region. The resultant film become more electron-doped with the Dirac point at -330 meV (Fig. 2f), which is likely due to creation of Se vacancies as a result of excess loss of Se during decomposition. [29] Note that this Dirac point energy is consistent with previous ARPES observation on the Bi2Se3 films using similar desorption parameters, [17] indicating similar degradation of Bi2Se3 films in previous studies. Therefore, although the direct thermal desorption can sublime the Se capping layer on Bi2Se3 surface, it cannot effectively remove the nanosize contaminants. The origin and composition of these nanometer-size contaminants is unknown in present study. Likely, they formed on top of Se capping layer when the sample was exposed to air and migrated towards Bi$_2$Se$_3$ surface during the Se desorption process, as illustrated in the cartoon in Fig. 2g. If this picture were correct, removing these surface contaminants, e.g. by ion sputtering, before Se desorption would result in clean Bi$_2$Se$_3$ surface. Based on this conjecture, we designed an improved decapping process.

Figure 3a shows a cartoon of the improved decapping process, i.e., a brief ion sputtering followed by thermal desorption of the Se capping layer. Using this method, the resultant surface is practically free of the nanosize contaminants. A typical STM image of such surface is shown in Fig. 3b. Merely 1% of this surface is covered with the contaminants, while the majority of the surface is atomically clean, as shown in a zoomed-in STM image (Fig. 3c) with atomic resolution. More importantly, the STS data taken over the clean area suggest that the Dirac point of surface state is at -145 meV, i.e., the Fermi level is ~55 meV below conduction band minimum. This indicates that our Bi$_2$Se$_3$ films are almost in the intrinsic semiconductor limit. Consistently, very low density of point defects is observed on such surface. This attests to the extremely high quality of our MBE Bi$_2$Se$_3$ films. Note that such clean decapping and high quality surface as shown in Fig. 3 are routinely obtained over multiple samples. More importantly, if the Se surface is not cleaned thoroughly, one could end up with inhomogeneous decapping, as shown in Fig. S4. Local STS measurements (Fig. S5) indicate that the surface contaminants donate electrons to Bi$_2$Se$_3$ film. With this improved Se decapping process, high quality Bi$_2$Se$_3$ thin film are now easily accessible to ex-situ characterization and/or device fabrications. Furthermore, such method with mild decapping temperature can be applied to other selenides or tellurides thin films, including topological insulators, Fe-based superconductors, etc.

Our systematic STM studies clearly demonstrate that even the inert surface of van der Waals bonded selenides would be contaminated by adsorption or oxidation if it is exposed to air without protection. Therefore, it is necessary to protect the pristine surface with Se capping layer. However, with the Se capping and direct thermal decapping, majority of the decapped surfaces are still covered with nanosize contaminants, which still prevents direct access of the pristine surface. The surface contaminants originally on top of amorphous Se layer are likely native oxides (SeO$_x$) due to exposure to ambient environment. Nominal physical adsorbates would be easily removed by thermal desorption. Previous X-ray photoemission spectroscopy (XPS) studies have demonstrated that SeO$_x$ form even on the surface of Bi$_2$Se$_3$ crystals when exposed to air, indicating the binding energy of SeO$_x$ is larger than that of Bi$_2$Se$_3$. [6] This is consistent with the persistence of the surface contaminants when the surface of Bi$_2$Se$_3$ film starts to decompose at high $T_d$ > 350 °C. (Fig. 2) Formation of native oxides is consistent with earlier XPS results from Hunger et al., [15] where trace amount of oxygen signal was observed on ZnSe film even with $T_d$ =390 °C. Our XPS studies of Se capping layer are consistent with this scenario. (See ESM Fig. S6) Note that our results do not exclude other possibilities, such as compounds of carbon. Further detailed surface spectroscopy studies are required to understand the exact nature of the contaminants. In all, the proper Se decapping temperature should be ~200-300 °C. Once the upper surface contamination layer on the amorphous Se was removed by ion sputtering, atomically clean $Bi_2Se_3$ surface could be obtained with $T_d$ as low as ~210 °C.

In conclusion, we have developed an improved Se decapping method to recover pristine $Bi_2Se_3$ surface. Our results demonstrated that extremely clean surface could be obtained with such method, which opens up a new paradigm of ex-situ surface characterization or device fabrication of high quality $Bi_2Se_3$ thin films. Furthermore, because of the low decapping temperature (~210 °C for $Bi_2Se_3$) such method might be applicable to other selenide materials, such single layer superconductor FeSe/STO [27], charge-density wave system $NbSe_2$, semiconductor ZnSe, photovoltaic $CuInSe_2$, etc. Beyond the Se family, one could also imagine similar recipes for oxide surfaces that are inert to amorphous Se.

**Methods:**

**MBE films**: High quality $Bi_2Se_3$ films with thickness of 100 quintuple layers (QLs) were grown on sapphire substrates (0001) following the recipe described elsewhere [17,30,31]. The amorphous Se was deposited on the $Bi_2Se_3$ film inside the growth chamber after the films were cooled to room temperature. The samples were transferred into the loadlock of our STM system via ambient environment after taken out from the dedicate MBE system.

**Decapping details**: The samples were introduced into a prep chamber with base pressure of $2\times10^{-10}$ mbar where the decapping was carried out. In the direct desorption process, the samples were heated up from room temperature to ~200 °C when the Se desorption was observed visually. Immediately after the desorption, we remove the samples from the heater allowing for a rapid cooldown. The entire annealing process takes about 5 minutes. Following the initial STM study, some of the samples were further studied after being re-annealed to even higher temperatures (up to ~350 °C). In the proper decapping method that we develop in this study, the samples were firstly sputtered with 500 eV argon ions which thoroughly cleans the upper surface of the amorphous Se while leaving the $Bi_2Se_3$ film, which is 100 nm down, intact. The sputtering current is about 1 µA and it usually lasts for ~10 minutes. Same annealing up to ~200 °C was applied subsequently to desorb the amorphous Se, which visually appears alike with the direct desorption process.

**STM/STS**: We used an Omicron LT-STM with base pressure of $1\times10^{-11}$ mbar. The STM/STS experiments are carried out at 78K, 48 K or 4.5 K using liquid (solid) nitrogen, or liquid helium. The differential conductance measurement was performed with the standard lock-in technique with amplifier gain $R_{gain}$ = 3 GΩ, modulation frequency $f$ = 455 Hz and amplitude $V_{mod}$ = 5 mV.

## Acknowledgements

We thank Yue Cao and Dan Dessau for helpful discussions. Thanks to Ryan Thorpe for assistance with the XPS experiments. J.D., W.B.W. and W.W. would like to acknowledge support by NSF DMR-0844807. M.B., N.K., M.S. and S.O. would like to acknowledge NSF DMR-0845464 and ONR N000141210456.

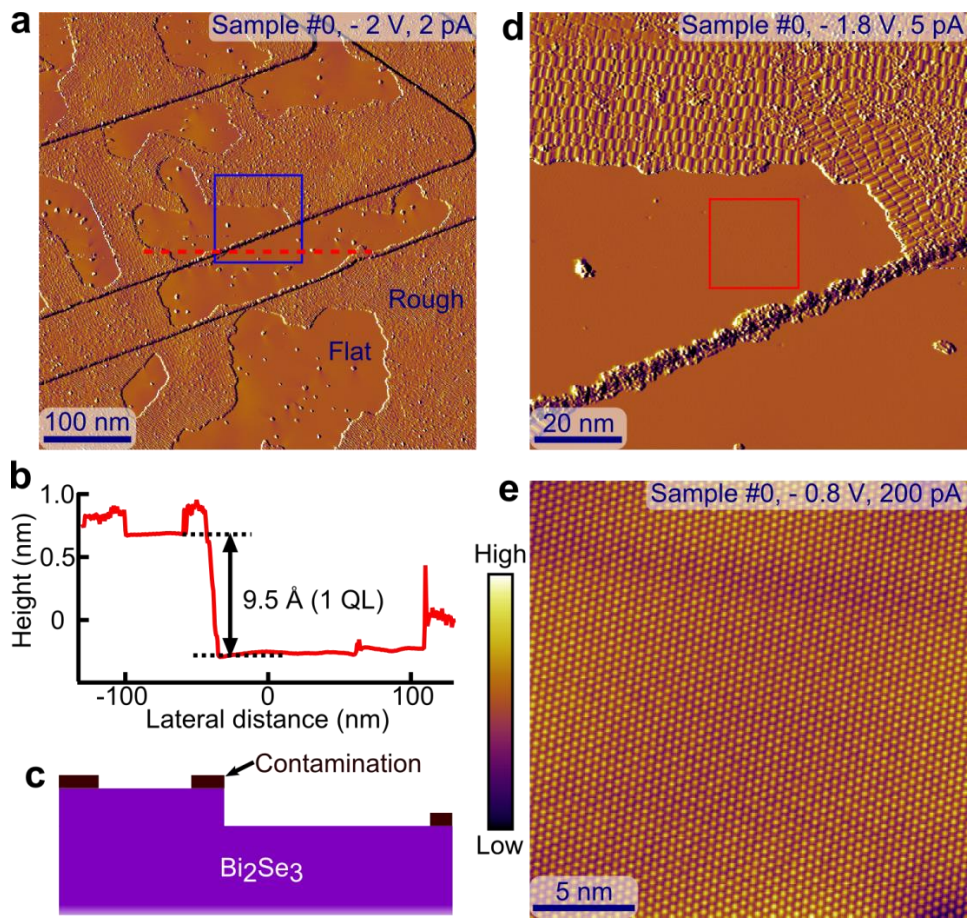

**Figure 1** Topographic images of bare Bi$_2$Se$_3$ film that was briefly exposed to air. (a) A large area (600 nm) scan showing large terraces along with straight step edges. (b) zoom-in of the area indicated by the blue box in (a) showing the detailed structures of the flat and rough areas. (c) profile along the dashed line in (a). (d) a diagram explaining the profile above. (e) zoom-in of the flat area in (b) inside the red box showing the triangular lattice of the top Se layer with lattice constant of roughly 4 Å. (a) and (b) are derivatives of the original height images. The tunneling parameters are specified inside each image.

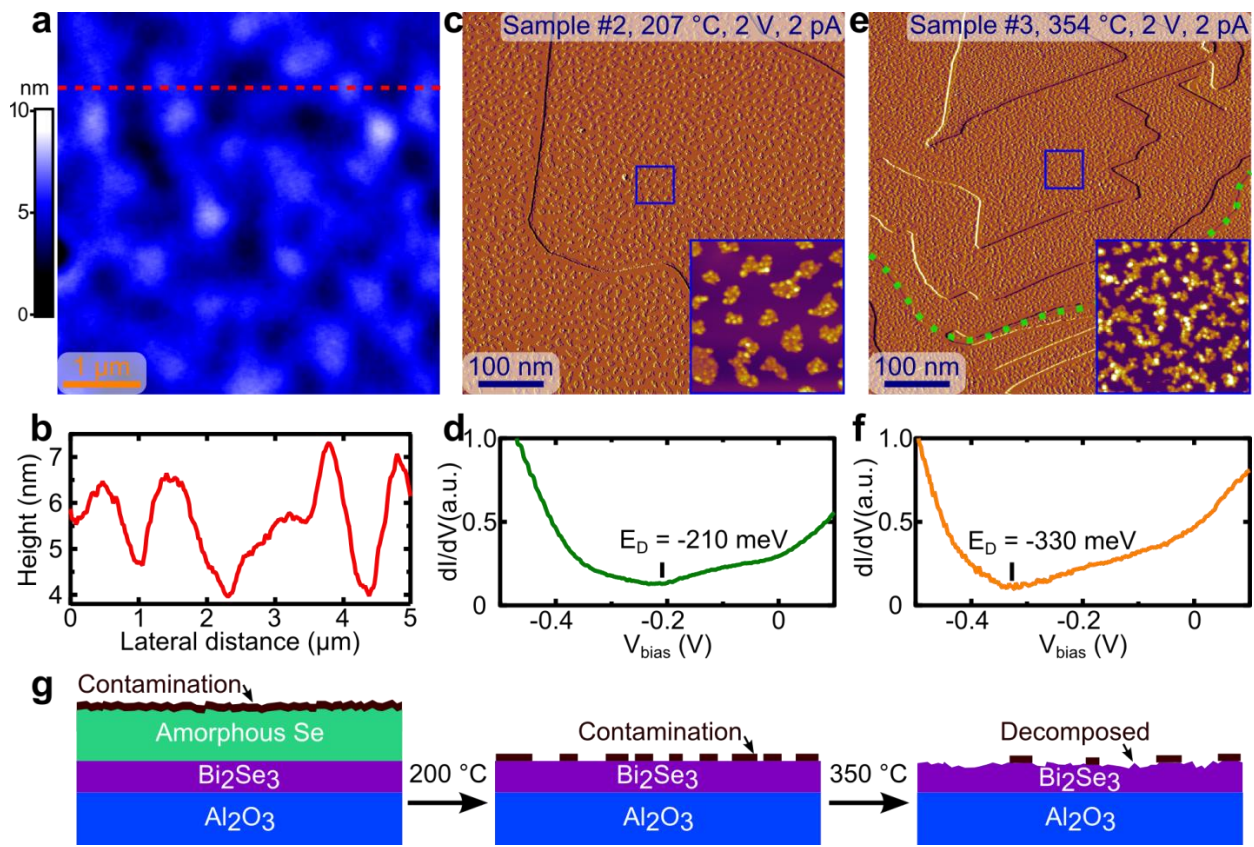

**Figure 2** Films that were capped with Se and decapped by direct annealing. (a) is an AFM image (5μm × 5μm) showing the morphology of the sample with Se capping. (b) is a height profile along the red dashed line. (c) and (e) are derivative images of two samples with direct decapping with annealing at 207 °C and 354 °C, respectively. The insets are detailed images of the areas inside the boxes. The green dashed line in (e) indicate a grain boundary. (d) and (f) are typical dI/dV spectra measured on dirt free regions of (c) and (e) (purple colored area in the inset images), respectively. (g) is a diagram explaining the direct depcapping process.

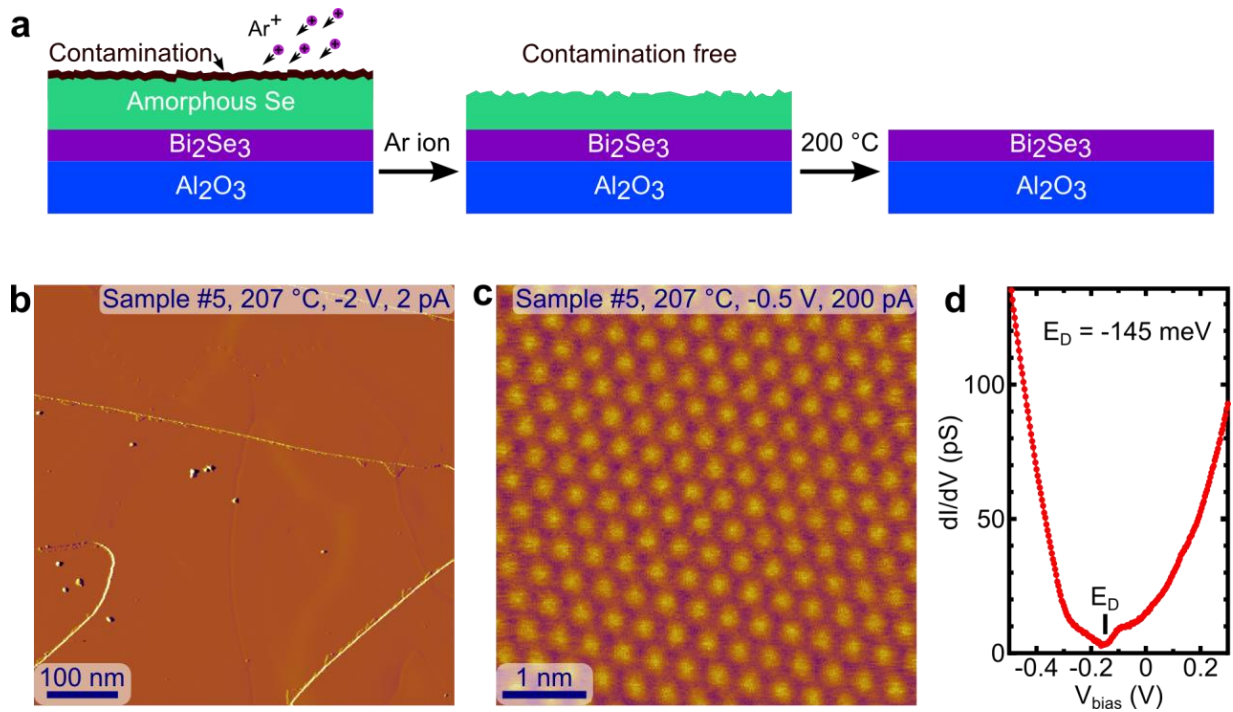

**Figure 3** A recipe for achieving ultra-clean decapping. (a) a diagram explaining the process of Ar ion sputtering followed by the mild annealing. (b) proper decapping yields extremely clean surface (derivative image). (c) atomically flat surface showing the triangular lattice of Se. (d) Spatially averaged spectroscopy measurement over the area shown in (c) showing the Dirac point of the Bi2Se3 film to be at -145 meV. Tunneling conditions for dI/dV measurement: Vset = -500 mV and Iset = 20 pA.

# Supplementary Material

# Restoring Pristine Bi$_2$Se$_3$ Surface with an Effective Se Decapping Process

Jixia Dai, Wenbo Wang, Matthew Brahlek, Nikesh Koirala, Maryam Salehi, Seongshik Oh, and Weida Wu

Department of Physics and Astronomy, Rutgers University, Piscataway NJ 08854, USA

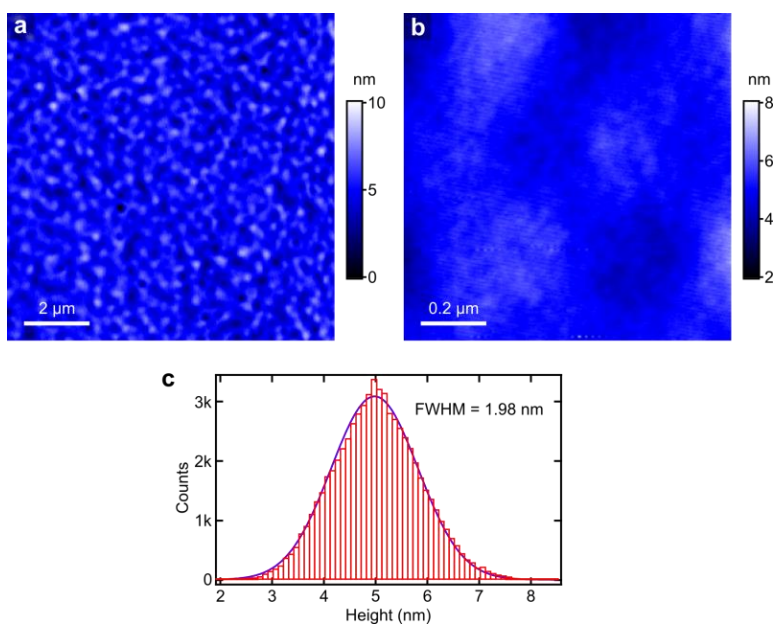

**Figure S-1.** Morphology of the Se capping layer on large (20 μm as shown in (a)) and small (2μm in (b)) scales. (c) is the histogram (red) of the image in (a) with a Gaussian fitting (purple) to the profile. The full width at half maximum (FWHM) of this distribution is about 2 nm, indicating that the Se capping is not atomically flat.

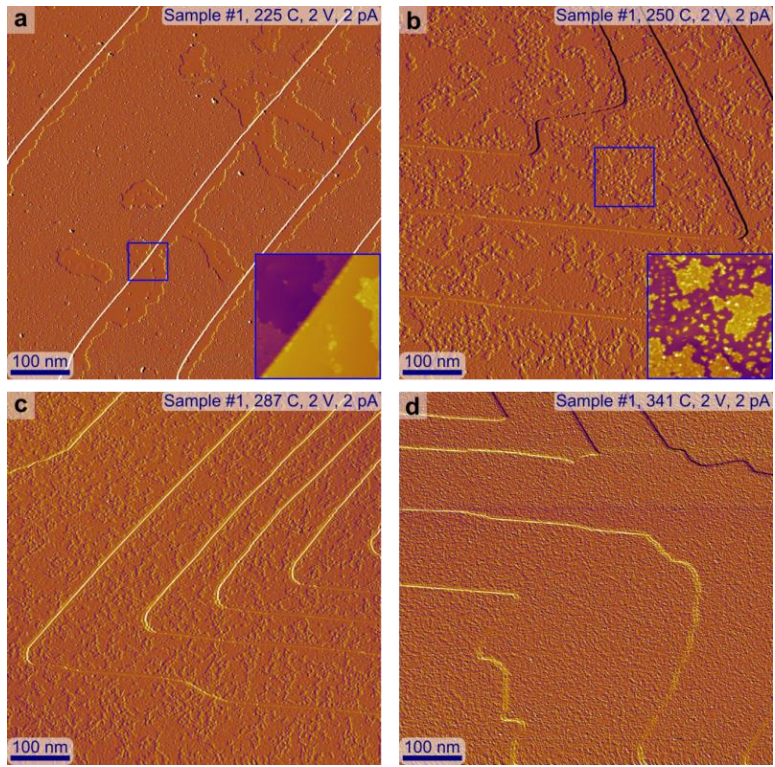

**Figure S-2.** Se capped Bi2Se3 film sequentially annealed to different temperatures. The surface contamination layer became progressively disordered while the terrace morphology kept unchanged until the highest temperature (341 °C in (d)) where the step edges became 'kinked'. Through the entire process the majority of the surface is atomically rough. The large scale images are derivative of the raw images.

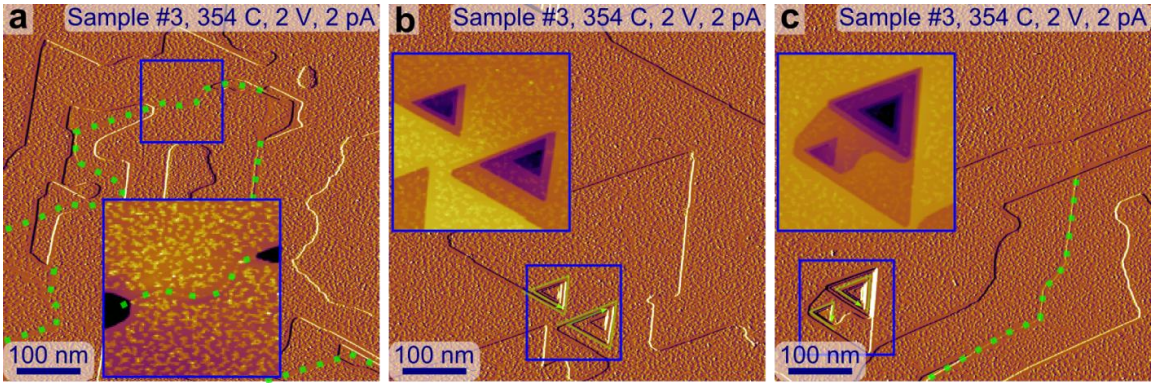

**Figure S-3** Morphology of overheated Bi2Se3 film. The images, including the one in Fig. 2(e), were taken on four different locations of the same sample. The triangular shaped holes in (b) and (c) are centered around screw dislocations (indicated by the lines with arrows) and the neighboring ones have opposite helicity (Burgers vector). The dashed lines indicate the grain boundaries. The insets are zoom-in height images of the area inside the small boxes.

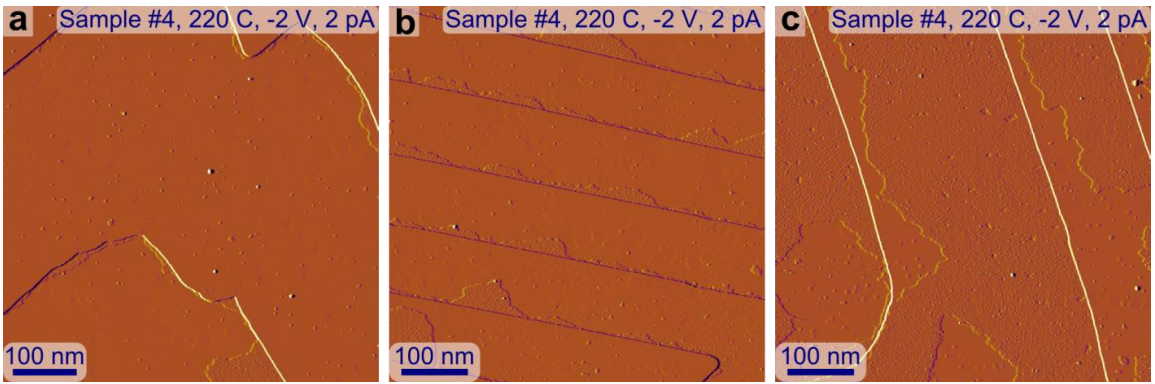

**Figure S-4** Insufficient sputtering causes inhomogeneous contamination coverage. (a) is taken at the center of the Ar ion sputtering, while (b) and (c) are taken at locations around 2 and 4 mm away from the sputtering center. (a) only shows some scattered contamination, while (b) has more continuous contamination along the step edges. In (c) the majority of the surface is covered with a contamination layer, very similar to the one shown in Fig. S1a where no sputtering was applied.

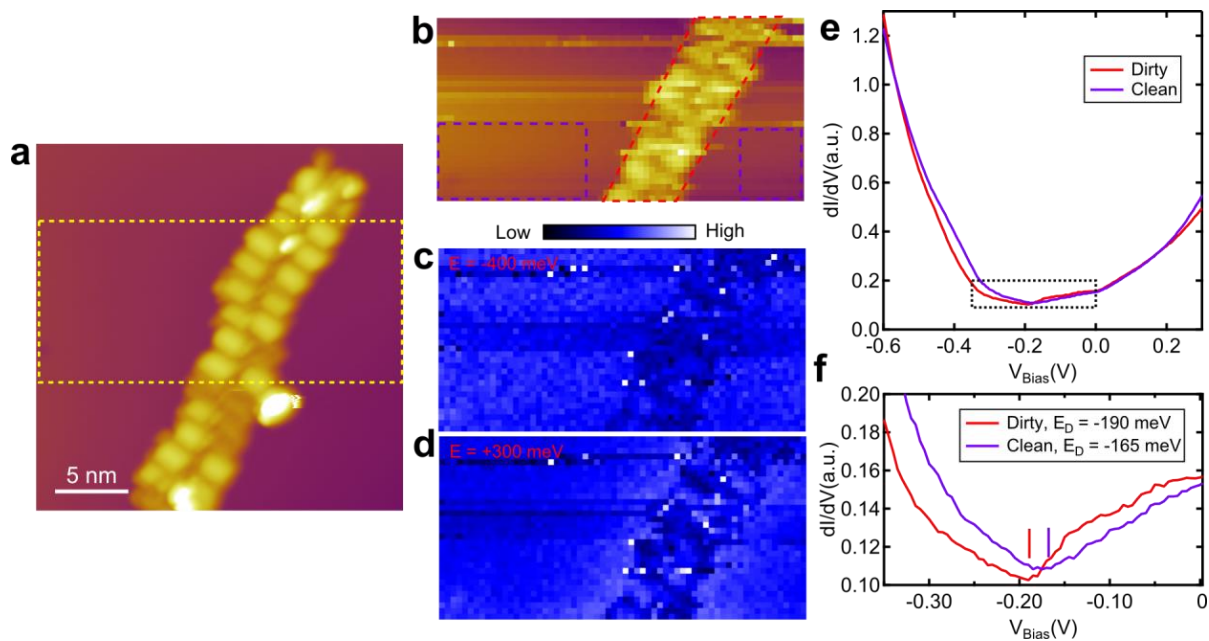

**Figure S-5** Scanning tunneling spectroscopy measurement on directly decapped sample. (a) Topography of an area with some dirt lying on top of $Bi_2Se_3$ surface. (b) Topographic image during the spectroscopy measurement shown in (c-f). (c) and (d) Spectroscopic images at -400 meV and +300 meV. (e) Averaged spectroscopy of the clean and dirty areas defined in (b). (f) Zoom in of (e) showing the shift of Dirac point.

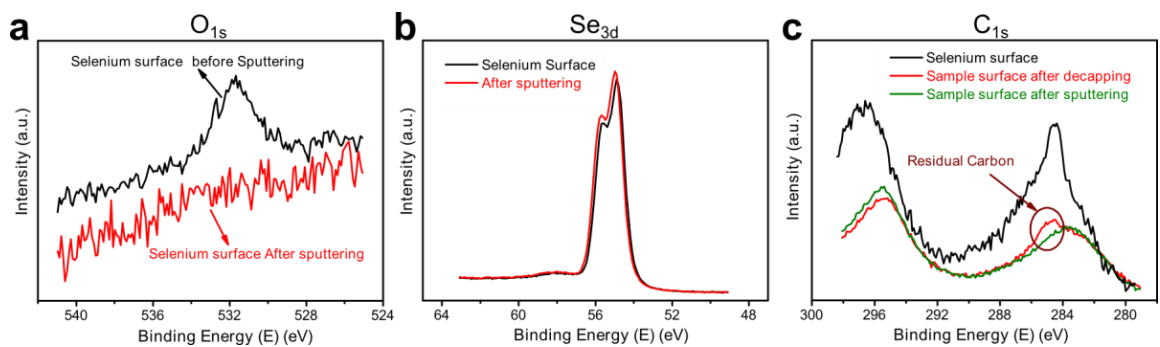

**Figure S-6** XPS measurements on capped and decapped films. (a) O 1s peak before and after ion sputtering. (b) Se 3d peak before and after ion sputtering. (c) C 1s peak before sputtering and decapping (black), which continues to show up in the directly decapped sample (red). An effective decapping with sputtering and heating is sufficient to get rid of the C 1s peak (green), which implies there is possibly carbon contamination. The double hump near 296 eV and 284 eV are Se Auger peaks.